\documentstyle[12pt]{article}

\def\be{\begin{equation}}
\def\ee{\end{equation}}
\def\bea{\begin{eqnarray}}
\def\eea{\end{eqnarray}}

\def\br{\begin{thebibliography}{abc}}
\def\er{\end{thebibliography}}

\def\a{\alpha}

\def\d{\delta}

\def\D{\Delta}

\def\L{\Lambda}

\def\bar#1{\overline{#1}}

\def\Hat#1{\rlap{\kern.10em$\widehat{\phantom G}$}#1}
\def\HAt#1{\rlap{\kern.05em$\widehat{\phantom G}$}#1}

\def\cAp#1{\rlap{\kern.1em$\widehat{\phantom{G\vrule height.8em}}$}#1{}}
\def\Cap#1{\rlap{\kern.05em$\widehat{\phantom{G\vrule height.8em}}$}#1{}}

\newcommand{\sect}[1]{\setcounter{equation}{0}\section{#1}}

\def\sxn#1{\bigskip\bigskip \sect{#1} \medskip}

\begin{document}

\thispagestyle{empty}
\setcounter{page}{0}

\begin{flushright}
PSU/TH/156 \\
UFIFT-HEP-95-16\\
SU-4240-598 \\
September 1995
\end{flushright}
\vspace*{15mm}
\centerline {\LARGE Chern-Simons Duality and the Quantum Hall Effect}
\vspace*{15mm}
\centerline {\large A. P. Balachandran$^{1}$, L. Chandar$^{1,2}$,
B. Sathiapalan$^{3}$}

\vspace*{5mm}
\centerline {\it $^{1}$ Department of Physics, Syracuse University,}
\centerline {\it Syracuse, NY 13244-1130, U.S.A.}
\centerline {\it $^{2}$ Department of Physics, University of Florida,}
\centerline {\it Gainesville, FL 32611, U.S.A.}
\centerline {\it $^{3}$  Department of Physics, Pennsylvania State University,}
\centerline {\it 120, Ridge View Drive, Dunmore, PA 18512, U.S.A.}

\vspace*{25mm}
\normalsize
\centerline {\bf Abstract}
\vspace*{5mm}
In previous work on the quantum Hall effect on an annulus, we used
$O(d,d;{\bf Z})$ duality transformations
on the action describing edge excitations to generate the Haldane hierarchy of
Hall conductivities.  Here we generate the corresponding hierarchy of ``bulk
actions'' which are associated with Chern-Simons (CS) theories, the connection
between the bulk
and edge arising from the requirement of anomaly cancellation.
We also find a duality transformation for the CS
theory exactly analogous to the $R\rightarrow \frac{1}{R}$ duality of the
scalar field theory at the edge.

\baselineskip=24pt
\setcounter{page}{1}
\newpage

\sxn{Introduction}

Chern Simons (CS) gauge theories 
are known to be particularly appropriate for describing the quantum Hall
effect (QHE) \cite{heff1,heff2}.
The low energy
effective action for the electromagnetic vector potential,
obtained by integrating out the electronic
degrees of freedom in a Hall system, is known to have a CS term.
The coefficient of this term is proportional to the Hall conductivity.
This fact is easily shown as follows.  Let us assume that the effective action
(apart from the usual Maxwell term) is
\be
S_{eff}[A] =-\frac{1}{2}\sigma _{XY}\int d^{3}x \epsilon
 ^{\mu\nu\lambda}A_{\mu}\partial
_{\nu}A_{\lambda} \label{One}
\ee
Then we get for the expectation value of the
current,
\be
< j ^{\mu} _{em} >_{A} =
- \frac {\delta}{\delta A _{\mu}} S_{eff}[A] =  \sigma _{XY}
 \epsilon ^{\mu \nu \rho}
\partial _{\nu} A_{\rho}
\ee
We thus see that there is a current in the X- direction
when there is an electric field in the Y- direction.  This is the
Hall effect, the Hall conductivity being $\sigma _{XY}$.


    In studying the Hall effect, we will be interested in CS theories
involving several vector potentials (one of which is the electromagnetic
field).  These comprise the ``statistical'' gauge fields and
the fields describing excitations in the bulk
\cite{heff1,heff2,zee,froh1,frad}.
The former are introduced for the purpose of changing the statistics
of the excitation fields in the action while the
latter represent collective degrees of freedom such as vortices
or other quasiparticles and can describe both bosons and fermions.

Experimentally, the Hall conductivity in certain systems is
quantized in integers or in certain
definite fractions \cite{qhe} corresponding respectively to integer and
fractional QHE (IQHE and FQHE). Several scenarios have been
proposed to explain such quantizations.
The hierarchy schemes of Haldane \cite{hald} and Jain \cite{jain} are perhaps
the most
attractive, in that they explain most of the observed experimental
fractions.  CS theories of the type mentioned above, involving
several vector potentials, lend themselves naturally to these
schemes \cite{zee,lee}.  The Jain scheme has already been written in this form
\cite{zee} while we will work out a similar description of the
Haldane scheme in this paper.

However, the CS action is not gauge invariant on a manifold with
a boundary (like an annulus), such a manifold being the appropriate geometry
for
a physical Hall sample.  One has to
include non-trivial dynamical degrees of freedom at the edge
\cite{froh1,wil,stone,renn,froh2} to restore gauge invariance.
In this way, one predicts the
existence of edge states, which nicely corroborates completely different
arguments showing the existence of chiral edge curents in a Hall sample
\cite{halperin}.  These states have been studied in detail by
many authors \cite{froh1,wil,stone,renn,froh2,edge,dim,dhe}.  In \cite{dhe},
we described them by a conformal field theory of massless chiral
scalar fields taking values on a torus.  The most general action for these
scalar fields contains a symmetric matrix $G_{ij}$ and an antisymmetric
matrix $B_{ij}$.  In \cite{dhe}, we saw that the Hall
conductivity depends on $G_{ij}$.  In this paper, we will show, in
detail, how
the anomaly cancellation
argument enables us to relate this matrix to a corresponding
matrix in the bulk CS theories.  Thus, once we implement the
hierarchy arguments in the CS theory, we have a rationale for
particular choices of this matrix made in \cite{dhe}.

Now, as in string theories having torally compactified spatial dimensions,
there are
certain duality transformations of the edge theory that leave the spectrum
invariant \cite{dua}.
These transformations change $G_{ij}$ and $B_{ij}$ in well-defined
ways and hence also change the Hall conductivity \cite{dhe}.
It was shown in \cite{dhe} that one can reproduce most of
 the conductivity fractions
of the Haldane and Jain schemes by means of these
generalized duality transformations.  The connection of the edge theory with
the CS theory in the bulk then suggests that similar
transformations can be implemented in the bulk CS theory also.
This conjecture turns out to be at least partly realizable in that one can
implement a duality transformation of the type $R \rightarrow 1/R$ in the bulk.
The demonstration of this result is a generalization of the one that has been
used in \cite{bus} to implement duality in scalar theories.  We think
that both this proof and the result are interesting
and could have implications in many other areas as well.

            This paper is organized as follows: In Section 2 we
describe how to implement the Haldane construction using CS
theories.  In Section 3 we describe the connection between the bulk and
edge actions.  Finally, in Section 4 we show how to implement duality in the CS
theory.

\sxn{The Haldane Hierarchy and CS Theory}

In this Section we would like to describe Haldane's construction
using CS gauge fields.  Let us first recall the physical arguments.
The Haldane
approach exploits the superfluid analogy and treats
the Hall fluid as a bosonic condensate.

We have a system with $N_{e}$ electrons per unit area in a magnetic
field of strength $B$.  The number of flux quanta per unit
area is $Be /2 \pi $ (in units where $\hbar = c =1$), which we denote
by $N_{\phi}$.  In the usual integer effect with one filled Landau level,
we have the equality
\be
N_{\phi} = N_{e} \label{iqhe1}
\ee
which tells that the degeneracy of the Landau levels is exactly equal to the
number of flux quanta piercing the Hall sample.  This equation is a consequence
of solving the Landau level problem for {\em fermions}.  The
incompressibility follows from the existence of a gap between
Landau levels and the fermionic nature of the electrons that
fixes the number of electrons that one can place in one Landau level.

There is another way to think of the same system.  According to
(\ref{iqhe1}) the number of flux quanta is equal to the number
of electrons.
It is therefore like
attaching one flux quantum to each electron.
The composite object behaves like a boson and can Bose condense.
The resulting superfluid is the Hall fluid.  The energy gap
follows from the usual arguments for superfluidity due to Feynman
\cite{feynman}, where he showed, using the bosonic nature of the condensate,
that the only low energy excitations are long wavelength density fluctuations.
However, in the Hall fluid, since the density is tied to the fixed external
magnetic field by (\ref{iqhe1}), there are no density fluctuations.
Thus we have no massless excitations (in the bulk) at all, that is, the fluid
is incompressible.

    A simple generalization of the above arguments can be applied to
a system that obeys
\be
N_{\phi}=m N_{e}\;\; , \; m\in 2{\bf Z}+1 .\label{fqhe1}
\ee
It can be interpreted as the attachment of
an odd number ($m$) of flux tubes to each electron.  Since
the composite will be bosonic, there will be bose condensation
and one can again invoke arguments from superfluidity.
Thus we have a new incompressible
state at the filling fraction $1/m$.  These are the Laughlin fractions.

Thus in the Haldane approach, the final
dynamical degrees of freedom are bosonic objects, a circumstance which suggests
that we rewrite the original action, which describes
fermions (electrons) in a magnetic field, in terms of a new set
of variables that describe these dynamical excitations.
Thus we will re-express the fermionic electron field in terms
of a bosonic field and a statistical gauge field.
As we shall see below, one can implement, in this way, the
ideas described in previous paragraphs,
in terms of a low energy effective field theory.

One can also proceed to generalize these ideas to get other filling
fractions.   The system admits Nielsen-Olesen vortices \cite{nielsen} as
excitations or quasiparticles.
As the magnetic field is changed, it is energetically
favourable for the excess or deficit magnetic field to organize itself as
 flux tubes threading vortices in the condensate, so that  quasiparticles
which
are one of these Nielsen-Olesen vortices are formed.  At a certain point
a large number of these quasiparticles form and condense, so
that we now have a finite number density of quasiparticles
and a new ground state is also created.

If one were to think of these quasiparticles or vortices as carrying a new form
of charge, then the gauge field to which they couple are in
fact the duals to the Goldstone (phase) mode of the
condensate \cite{zee,lee}.  The way one shows this point \cite{zee,lee} is by
noticing that if the electron current is represented in a dual representation
using a one-form, then the ``electric field'' associated to this one-form has a
behaviour, outside a vortex, identical to that of a usual electric field
outside an ordinary electric charge.

Thus the flux
quanta, in this dual representation, are the electrons
themselves.  The quasiparticles are bosons.  Clearly, their bosonic nature will
be maintained if an even number (say $2p_{1}$) of dual flux quanta (that is,
electrons)
get attached to each of these quasiparticles.  These statements can be
summarized by the following two equations:
\be
N_{\phi} = m N_{e} + N^{(1)}                  \label{ad}
\ee
\be
N_{e} = 2 |p_{1} N^{(1)}|  \label{mod}
\ee
Here $N^{(1)}$ is the number density of quasiparticles.  Unlike $N_e$, it can
have either sign
depending on whether the
associated fluxons point in a direction parallel or antiparallel to
the original magnetic field.

If we make $p_1$ negative whenever $N^1$ is, then we can omit the
modulus sign in equation (\ref{mod}).  In that case, we can solve these
equations to get the filling fraction:
\be
\frac {N_{e}}{N_{\phi}} = \frac{1}{m+ \frac{1}{2p_{1}}}
\ee
This is the second level of the hierarchy.

We can next imagine that there are new quasiparticle excitations
over the ground state as we increase the magnetic field further.
These new quasiparticles can have ``flux tubes" attached to them and
can in turn condense.  Now the new flux quanta are dual representations
of the quasiparticles of the first level.  This process can be iterated
as many times as one wants and it generates a series of filling fractions.
The equations describing this process are the hierarchy equations of
\cite{zee}:
\be      \label{eq:hier1}
N_{\phi} = m N_{e} + N^{(1)} ,
\ee
\be                      \label{eq:hier2}
N_{e} = 2 |p_{1} N^{(1)}| + N^{(2)}   ,
\ee
\be
N^{(1)} = 2 |p_{2} N^{(2)}| + N^{(3)}   ,\label{eq:hier3}
\ee
\[
  .......
\]
In equations (\ref{eq:hier2}), as in (\ref{ad}), the
quasiparticle density $N^{(2)}$ can be less than zero, but should still be such
that $N_e$ itself does not become negative.

We can in fact choose to omit the modulus signs in these equations if we allow
the $p_i$'s also to be less than zero whenever the $N^i$'s are, so that their
product themselves are always non-negative.  We shall assume that this
is done in the
following, where we implement these ideas using CS fields.  The basic
techniques are described in \cite{zee}.

Following \cite{zee}, we describe the electron by a scalar field
coupled to a
statistical gauge field, $a_\mu$. Furthermore if this bosonic order parameter
develops an expectation value, then we have a massless Goldstone boson
$\eta$ -
the phase of the original scalar field. It fulfills the equation $\partial
_{\mu}\partial ^{\mu} \eta =0$, being massless.  The electron current
$\partial ^\mu
\eta$ can be represented by a dual vector field $\alpha_\mu$ defined by
$\partial ^\mu \eta = \epsilon^{\mu \nu \lambda} \partial_\nu
\alpha_\lambda$. The field equation of $\eta$ turns into an identity in this
dual representation.  We can also implement a minimal coupling to the external
electromagnetic vector potential $A_\mu$. The action thus far is
\begin{eqnarray}
&&\int _{(D\backslash H)\times {\bf R^{1}}}d^{3}x\; [-eJ^\mu ( A_\mu -
a_\mu )-
\frac{e^2}{4\pi}\epsilon ^{\mu \nu \lambda}a_\mu \partial_\nu a_\lambda ] ,
\nonumber\\
&&J^\mu =\epsilon ^{\mu \nu \lambda} \partial_\nu \alpha_\lambda
\label{4.1}
\end{eqnarray}
\[ D\backslash H \equiv \mbox{ Disk D with a hole H removed (or an
annulus)} \]
where the last term is an abelian CS term for the statistical gauge field
$a_\mu$ and ${\bf R^{1}}$ accounts for time.  Its coeffecient has been
chosen to ensure that it converts the
boson to a fermion as may be seen in the following way: On varying (\ref{4.1})
with respect to $\alpha_\mu$, we get
\be
\epsilon ^{\mu\nu\lambda}\partial _{\nu}A_{\lambda}= \epsilon
^{\mu\nu\lambda}\partial _{\nu}a_{\lambda} .\label{cancellation}
\ee
 On varying with respect to $a_\mu$, we get
\begin{equation}
\epsilon ^{\mu\nu\lambda}\partial_\nu \alpha_\lambda = \frac{e}{2\pi} \epsilon
^{\mu\nu\lambda}\partial_\nu a_\lambda .
\label{4.2}
\end{equation}
so that
\begin{equation}
\epsilon ^{\mu\nu\lambda}\partial_\nu \alpha_\lambda = \frac{e}{2\pi} \epsilon
^{\mu\nu\lambda}\partial_\nu A_\lambda .\label{4.3}
\end{equation}

This equation relates the number density $J^0  =N_e $ of electrons to the
number
density $N_\phi =\frac{e}{2\pi}B$ of flux quanta $\frac{2\pi}{e}$. In fact it
says that $N_e = N_\phi$.  Thus there is one flux quantum per electron which
converts the latter to a fermion.

The filling fraction $\nu$ is 1 for (\ref{4.1}) since $N_{\phi}=N_e$. It thus
describes the
IQHE (see (\ref{iqhe1})). We can also eliminate $\alpha$ and
$a$ to get an effective action dependent only on the electromagnetic gauge
field.  Thus the electromagnetic current $-eJ^{\mu}$ of (\ref{4.1}) is equal to
$-\frac{e^{2}}{2\pi}\epsilon ^{\mu\nu\lambda}\partial _{\nu}A_{\lambda}$ by
(\ref{4.3}) and this current is reproduced by
\begin{equation}
S= -\frac{e^2}{4\pi} \int _{M}d^{3}x\;\epsilon^{\mu \nu \lambda} A_\mu
\partial_\nu A_\lambda ,\label{4.4}
\end{equation}
\[
M= (D \backslash H) \times {\bf R^{1}} .
\]

This is the electromagnetic CS term ( and a signature of the Hall effect )
for the Hall conductivity $\sigma_H =\frac{e^2}{2\pi}$.

One can immediately generalize (\ref{4.1}) to obtain the Laughlin fractions
by
changing the coefficient $\frac{e^2}{4\pi}$ to $\frac{e^2}{4\pi m}$ with
$m$ odd: \be
S^{(0)}= \int _{M}d^{3}x\; [-eJ^\mu ( A_\mu - a_\mu )-
\frac{e^2}{4\pi m}\epsilon ^{\mu \nu \lambda}a_\mu \partial_\nu a_\lambda ]
, \;\; m\in 2{\bf Z}+1 .\label{4.1tr}
\ee
This changes (\ref{4.2},\ref{4.3}) to
\begin{equation}
\epsilon ^{\mu\nu\lambda}\partial_\nu \alpha_\lambda = \frac{e}{2\pi m}
\epsilon
^{\mu\nu\lambda}\partial_\nu a_\lambda ,\label{4.5}
\end{equation}
\begin{equation}
\epsilon ^{\mu\nu\lambda}\partial_\nu \alpha_\lambda = \frac{e}{2\pi m}
\epsilon ^{\mu\nu\lambda}\partial_\nu A_\lambda .\label{4.6}
\end{equation}
Equation  (\ref{4.5}) says that $N_e = \frac{1}{m} N_\phi$.  Since $m$ is
odd, this
is the same as (\ref{fqhe1}) and therefore implies that the composite is
bosonic, as it should be for this description of the electron to be consistent.
The filling fraction now is $\nu = \frac{1}{m}$ while (\ref{4.4}) is changed to
\begin{equation}
\bar{S}^{(0)}= -\frac{e^2}{4\pi m}\int _{M}d^{3}x\; \epsilon^{\mu \nu
\lambda} A_\mu
 \partial_\nu
A_\lambda
\label{4.7}
\end{equation}
This is the CS action giving the first level of the Haldane hierarchy.

Next, we modify (\ref{4.1tr}) by adding a coupling of the quasiparticle current
$J^{(1)\mu}$ to the gauge field $\alpha_\mu$. Thus we have the action
\begin{equation}
\int _{M}d^{3}x\;[-eJ^\mu ( A_\mu - a_\mu )- \frac{e^2}{4\pi
m}\epsilon^{\mu \nu
 \lambda} a_\mu \partial_\nu a_\lambda + 2 \pi J^{(1)\mu}\alpha_\mu ]
\label{4.11}
\end{equation}

The choice of the coefficient $2\pi$ in the last term can be motivated as
follows.  Suppose that there is a vortex localised at $z$ so that
$J^{(1)0}(x)=\delta ^{2}(x-z)$ while the electron density $J^{(0)}$ is some
smooth function.  Then since $J^{(0)}=\frac{e}{2\pi m}\epsilon ^{0ij}\partial
_{i}a_{j}$ by equations of motion, $\epsilon ^{0ij}\partial _i a_j $ is also
smooth.  Now variation of $\alpha$ gives $\frac{2\pi}{e}J^{(1)0}=\epsilon
^{0ij}(\partial _i A_j +\partial _i a_j )$ so that the magnetic flux attached
to the vortex is the flux quantum $\frac{2\pi}{e}$.  As this is the unit of
magnetic flux we want to attach to the vortex, the choice of $2\pi$ is seen to
be correct.

Suppose next that the quasiparticles condense.  Then we can write
$J^{(1)\mu} = \partial^\mu \eta^{(1)}$
where $\eta^{(1)}$ is the Goldstone boson phase degree of freedom. As before,
$\eta^{(1)}$ being massless and hence $\partial _\mu \partial ^\mu \eta
^{(1)}=0$, one can write a dual version of the current by
defining a field $\beta_\mu$ according to
\begin{equation}
J^{(1)\mu}= \partial^\mu \eta^{(1)} = \epsilon^{\mu \nu \lambda} \partial_\nu
\beta_\lambda .\label{4.12}
\end{equation}
We can also introduce a statistical gauge field $b_\mu$ and attach flux tubes
of $b$ to the quasiparticle.
Since the quasiparticles correspond to vortices which are assumed to be
bosonic, here we attach an even number of the elementary $b$ flux tubes to
each vortex to
preserve the bosonic nature.  Bearing this in mind, we add some more CS terms
to (\ref{4.11}) to get
\begin{equation}
S^{(1)}= \int _{M}[-e ( A - a )d\alpha -
 \frac{e^2}{4\pi m}
a da + 2 \pi \alpha d\beta - e bd \beta - \frac{e^2}{4\pi (2 p_1)} b db ]
,\;\; m\in 2{\bf Z}+1,\;\; p_i \in {\bf Z}.\label{4.13}
\end{equation}
[Here, we have used the form notation to save writing the antisymmetric symbol
repeatedly.  A symbol $\xi =A,\alpha ,\beta ,a$ or $b$ now denotes the
one-form $\xi _\mu dx^\mu $.]

The equations of motion from (\ref{4.13}) are
\begin{eqnarray}
&& \frac{e}{2\pi}dA=\frac{e}{2\pi}da +d\beta ,\nonumber\\
&& md\alpha =\frac{e}{2\pi}da ,\nonumber\\
&& d\alpha = \frac{e}{2\pi}db ,\nonumber\\
&& d\beta = -\frac{e}{2 \pi (2 p_1)} db
\label{4.14}
\end{eqnarray}
The equations for $\alpha$ and $\beta$ here are seen to be precisely the
hierarchy equations (\ref{eq:hier2}) and (\ref{eq:hier3}) [with $N^{(2)}=0$] on
eliminating $a$ and $b$.

Now these equations for $\alpha$ and $\beta$ are reproduced also by
\begin{eqnarray}
\bar{S}^{(1)}& = &  \int _{M}[-e A d\alpha + \pi m
 \alpha d
\alpha + 2\pi \alpha d\beta + \pi (2p_1 ) \beta d
\beta ]\label{4.15} \\
& = & \int _{M}[-e A d\alpha + \pi ( \alpha \;\; \beta
 )\left(
\begin{array}{cc} m & 1 \\
1 & 2p_1 \end{array}\right)\left( \begin{array}{c} d\alpha \\
d\beta \end{array} \right ) ].\label{4.16}
\end{eqnarray}
We have here used matrix notation to display the form of the ``metric" in the
CS
theory.

The generalization to higher levels is as follows: Introduce $d$ vector
fields $\alpha_I$; $I=1, \cdots, d$. [In the above
example, $d=2$, $\alpha_1 = \alpha$, $\alpha_2 = \beta$.]  Then consider
the Lagrangian form
\begin{equation}
{\cal L} = -e A d\alpha_1 + \pi \alpha_I K^{IJ} d\alpha_J
\label{4.17}
\end{equation}
with
\begin{eqnarray}
&&\alpha _I =\alpha _{I\mu}dx^{\mu}, \nonumber\\
&&K^{IJ} = \left( \begin{array}{cccccc} m & 1 & 0& \cdot & \cdot & \cdot \\
1 & 2p_1 & 1 & 0 & 0 & \cdot \\
0 & 1 & 2p_2 & 1 & 0 & \cdot \\
\cdot & 0 & 1 & 2p_3 & 1 & \cdot \\
\cdot & \cdot & \cdot & \cdot & \cdot& \cdot \end{array} \right)  .
\label{4.18}
\end{eqnarray}
The equation of motion for $\alpha_1$ gives
\begin{equation}
e dA =2\pi K^{1J} d\alpha_J
\label{4.19}
\end{equation}
while the equations of motion for the remaining $\alpha _{I}$'s give
\be
K^{IJ}d\alpha _{J}=0 \mbox{ for }I\neq 1.\label{ineq}
\ee
These are the hierarchy equations. We can solve for the $d\alpha_I$'s:
\begin{equation}
d\alpha_I = \frac{e}{2\pi}  (K ^{-1})_{I1} dA  .
\label{4.20}
\end{equation}
Substitute back into (\ref{4.18}) to get
\begin{equation}
\bar{\cal L} = -\frac{e^2}{4 \pi} A (K^{-1})_{11} dA  .
\label{4.21}
\end{equation}
This is the CS Lagrangian form that gives rise to the Haldane hierarchy.   Its
filling fraction $\nu$ is just $(K^{-1})_{11}$ where $K^{IJ}$ is given
by (\ref{4.18}).  $\nu$ is in fact the continued fraction obtained in the
Haldane hierarchy:
\be
\nu =\frac{1}{m-\frac{1}{2p_1 -\frac{1}{2p_2 -\frac{1}{2p_3 -\ldots  }}}}   .
\label{hierarh}
\ee
We can prove (\ref{hierarh}) easily.  Let
\be
\Delta (\xi _1 ,\xi _2 ,\ldots \xi _n ) =\mbox{det} \left[ \begin{array}{ccccc}
\xi _1 & 1& 0 & \cdot & \cdot \\
1 &\xi _2 & 1 & 0 & \cdot \\
0 & 1 & \cdot & \cdot & \cdot \\
\cdot &\cdot & \cdot & \cdot &\cdot \\
\cdot & \cdot & \cdot & 1& \xi _n \end{array} \right] .\label{deter}
\ee
Then
\be
\Delta (\xi _1 ,\xi _2 ,\ldots \xi _n )= \xi _1 \Delta (\xi _2 ,
\ldots \xi _n ) -\Delta (\xi _3,\ldots \xi _n ) ,\label{determ}
\ee
and
\be
\nu =\frac{\Delta (2p_1 ,2p_2 ,\ldots 2p_n )}{\Delta
(m,2p_1 ,\ldots 2p_n )}. \label{determi}
\ee
We get (\ref{hierarh}) from (\ref{determ}) and (\ref{determi}).

\sxn{Anomaly Cancellation and the Bulk-Edge Connection}
    In this section we will show that the requirement of gauge
invariance forces the ``metric" $K^{IJ}$ introduced in the
previous section to be the same as the inverse of the target space ``metric''
$G_{IJ}$
of the scalar theory describing the edge excitations.  In our
previous work \cite{dhe} on edge excitations, we had assumed
the form (2.25) for $(G^{-1})^{IJ}$.  The results
of this and the previous section provide the necessary motivation
for this assumption.
Let us consider the CS action

\be    \label{cs}
S = \frac{1}{2} \int _{{\tilde M}} \a d \a
\ee
without any electromagnetic coupling.
If ${\tilde M}$  has a closed (compact and boundaryless) spatial slice, this
action has the gauge invariance
\be
\a \rightarrow \a + d \L        \label{gtr}
\ee
If ${\tilde M}$ is a manifold such as $M$ where the spatial slice
$\Sigma$ has a boundary like an annulus $D\backslash H$, then the
gauge variation results in a surface term.  For $M=D\backslash H \times
{\bf R^1}$, we have, for the variation of the action,
\be    \label{st}
\d S = \frac{1}{2} \int _{\partial D \times R^{1}}\L d \a -\frac{1}{2}\int
_{\partial H \times R^{1}}\L d\a ,
\ee
where as usual we assume that $\Lambda $ vanishes in the infinite past
and future.
One can recover gauge invariance at the boundary by adding to the
action the following two-dimensional action containing a new
scalar field $\phi$ :
\be
\D S = -\frac{1}{2}\int _{\partial M} d\phi \a +\frac{1}{4}
\int _{\partial M}d^{2}x (\tilde{D}_{\mu}\phi )(\tilde{D}^{\mu}\phi )
\label{ba}
\ee
Here the gauge transformation law for $\phi$ is
\be
\phi \rightarrow \phi - \L     \label{phitr}
\ee
so that $\tilde{D}_{\mu}\phi$ is $\partial _{\mu}\phi +\a _{\mu}$.  [The
coefficient
$\frac{1}{4}$ outside the kinetic energy term in (\ref{ba}) is determined by
requiring that the edge current be chiral, that is, that we can impose the
following condition consistently with the equations of motion:
\be
\tilde{D}_{-}\phi \equiv (\tilde{D}_{0}-\tilde{D}_{\theta})\phi =0 \label{coef}
\ee (see \cite{dhe}).]

The combined action $S+\D S$ is gauge invariant.

A more formal way of justifying the above procedure to recover
gauge invariance is to first
look at the generators of the ``edge'' gauge transformations in the absence of
the scalar field action at tbe boundary.

The operator that generates the transformation (\ref{gtr})
at a fixed time
with $\Lambda
|_{\partial D} \neq 0$ and $\Lambda |_{\partial H}=0$ ($\Lambda $ being a
function on the annulus $D\backslash H$, the choice $\Lambda |_{\partial H}=0$
being made for simplicity) is
\be
Q(\Lambda ):= \int _{D\backslash H} d\Lambda \alpha .\label{Chargel}
\ee
The algebra generated by these operators is specified by (\cite{csb})
\be
{[} Q(\Lambda ),Q(\Lambda ') ]=-i\int _{\partial D}\Lambda d\Lambda '
.\label{KM}
\ee
If one tries to impose the gauge invariance condition
 $Q(\Lambda )|\cdot \rangle =0$ on physical
states $|\cdot \rangle$, one is led to a contradiction because the commutator
of two $Q$'s acting on a (physical) state would also have to vanish,
whereas (\ref{KM}) specifies the value of this commutator to be a
non-zero $c$-number.

However,
if we now augment this action by the above action (\ref{ba}) describing
new degrees of freedom at the boundary, the generators of the ``edge'' gauge
transformations get modified.  The modification is by the terms

\be
q(\Lambda )= \int _{\partial D} \Lambda (\Pi _{\phi}-\frac{1}{2}\phi '),
\label{chargel}
\ee
where $\Pi _{\phi}:=\frac{1}{2}(D_{0}\phi +A_{\theta } )$, is the
canonical momentum conjugate to $\phi$ and obeys the usual commutation
relations.

$q(\Lambda )$ generates the transformations
\begin{eqnarray}
\phi &\rightarrow & \phi -\Lambda \nonumber\\
\Pi _{\phi} &\rightarrow & \Pi _{\phi} +\frac{1}{2}\partial _{\theta }\Lambda
\label{gtrl}
\end{eqnarray}
The algebra generated by the $q(\Lambda )$'s is given by
\be
{[} q(\Lambda ),q(\Lambda ')] =i\int _{\partial D} \Lambda d\Lambda '
.\label{km}
\ee
Thus the new
generators $\tilde{Q}(\Lambda ):= Q(\Lambda )+q(\Lambda )$ now commute amongst
themselves and can be chosen to annihilate the physical states.

Let us now attempt to couple electromagnetism to the action $S$ in (\ref{cs}).
  $*d \a$ represents
a current so that the obvious coupling is
\be
S^{1} =- q\int _{M} A d \a  \label{min}
\ee
Here $A$ is a background electromagnetic field.

However, we run into a problem when we consider the equation
of motion implied by $S + S^{1}$.
On varying with respect to  $\a $, the equation of motion that we get in the
bulk is
\be \label{eqnbul}
d\a = qdA
\ee
while on the boundary, it is
\be \label{eqnbou}
\frac{1}{2}\a = qA    .
\ee
(\ref{eqnbul}) and (\ref{eqnbou}) are incompatible. (\ref{eqnbou})
implies a relation between the values of the field strengths of
$\a$ and $A$ on the boundary that differs by a factor of two
from that implied by (\ref{eqnbul}) in the bulk whereas by continuity
they should be equal.

There is, however, the following simple modification of the minimal
coupling (\ref{min})
that gives a consistent set of equations.  Consider the action
\be  \label{modmin}
S^{2} = -\frac{1}{2}\int q(Ad\a + \a dA)
\ee
With this action, the boundary equation (\ref{eqnbou}) is modified to
\be   \label{meqnbou}
\a =qA   .
\ee
 Thus (\ref{eqnbul}) and (\ref{meqnbou}) together say that
 $\a = qA$ everywhere classically, up to gauge transformations that
vanish on the boundary.
Gauge transformations  that do not vanish on the boundary
and consistent with the equations of motion have the form
\be                 \label{ngtr}
\a \rightarrow \a + qd\L  \, ,\, A \rightarrow A + d \L
\ee

But
while we have achieved consistency of the equations of motion
at the edge and in the bulk, the action $S+ S^{2}$ given
by (\ref{cs}) and (\ref{modmin}) is no longer gauge
invariant under (\ref{ngtr}).
This is very similar to what
happens at the edge \cite{dhe}, where gauge invariance and chirality
are incompatible with the equations of motion. The cure
there (see \cite{dhe}and references therein),
was to introduce a coupling to the bulk.  Similarly,
here, the cure is to couple to degrees of freedom living only at the
boundary, just as was done in the beginning of this section for the action
(\ref{cs}) (see (\ref{cs})-(\ref{ba})).
Thus we need to introduce a
scalar field $\phi $ with a boundary action of the form
\be  \label{ba1}
\D S ^{2} = \frac{q}{2}\int _{\partial M} d\phi A +\frac{1}{4}\int _{\partial
M} d^2 x (D_{\mu}\phi )^2 ,
\ee
\[ D_{\mu}\phi =\partial _{\mu}\phi -qA_{\mu} \]
to maintain invariance under (\ref{ngtr}), namely the
 electromagnetic gauge transformations.
Here $\phi $  transforms under
 (\ref{ngtr}) in the following way:
\be  \label{U1em}
\phi  \rightarrow \phi + q\L
\ee
[Once again, we can justify this addition by noting as before that with this
addition, the generators of the ``edge'' gauge transformations can be
required to annihilate the states.]

The full action ${\cal S}=S + S^{2} + \D S^{2} $ is thus gauge invariant
under the electromagnetic gauge transformations.  It is also easy to see that
it gives
equations of motion in the bulk and the boundary that are compatible with each
other.
The final action is thus
\be    \label{fina}
{\cal S}=\int _{D\backslash H \times R^{1}} \{ \frac{1}{2}\a d \a -
\frac{q}{2}(A d \a + \a dA) \} +\frac{q}{2}
\int_{\partial D \times R^{1}} d\phi A   +\frac{1}{4}\int _{\partial D\times
R^1}d^2 x (D_{\mu}\phi )^2
\ee

   Let us summarize what we have done with one CS field before we
generalize to the case of $d$ fields.  We began with a CS action
for a gauge field $\a$, where $d\a $ represents the current of
electrons or quasiparticles.
We then introduced a coupling to a background
electromagnetic field.
 Naively,
 this action has a gauge invariance  even without
introducing any edge degrees of freedom.
However there
is an inconsistency between the bulk
and the boundary equations.  When the naive
coupling is modified to
restore consistency, the action is no longer gauge invariant.
The solution is
to introduce a scalar degree of freedom at the edge. The final
action is then given by (\ref{fina}).

It is now straightforward to extend this to the case with
$d$ CS fields and the action
\be \label{dcs}
S = \pi K^{IJ} \int _{M}\a _{I} d \a _{J}
\ee
We have introduced
the ``metric" $K^{IJ}$ that we had in the last section.
  This theory has $d$ U(1) gauge invariances:
\be \label{dgtr}
\a _{I} \rightarrow \a _{I} + d \L _{I}
\ee

We now introduce $d$ background gauge fields $A^{I}$, one of which
represents the physical electromagnetic field and the rest
are fictitious.  They can be used, for instance,
to calculate correlations
between the different quasiparticle currents. Thus once we
integrate out the quasiparticles from the theory, the resultant
action will depend on these gauge fields.  Functional differentiation
with respect to these fields then gives the correlators of the currents
(the connected Green functions).
They are, thus, a
means of keeping track of the information in the original action
after integrating out the $\a $ exactly, much as the ``sources"
of conventional field theory. Following the earlier procedure as earlier of
first
introducing a coupling as in (\ref{modmin}) and then introducing edge scalar
 fields
for restoring gauge invariance, the final action becomes
\begin{eqnarray}
&&{\cal S}= \int _{M} \{ -\frac{1}{2}(A^{I} d \a _{I} + \a _{I} d A ^{I}) + \pi
K^{IJ} \a _{I} d \a _{J} \} +\int _{\partial M}
 \frac{1}{4\pi} (K^{-1})_{IJ} \phi ^{I} A ^{J} \nonumber\\
&&+\frac{1}{8\pi} \int _{\partial M} (K^{-1})_{IJ} D _{\mu} \phi ^{I}
D^{\mu} \phi ^{J}
\end{eqnarray}
As in (\ref{ba}), here too the coefficient of the kinetic term is fixed by
requiring consistency between the chirality of the edge currents (\ref{coef})
and the equations of motion \cite{dhe}.

We can also specialize to the case where only the
electromagnetic background is non-zero.  Then we can
set $A^{I} = q^{I} A _{em}$ and get the expression
for the Hall conductivity used in \cite{dhe}.  The expression
used in the last section is obtained by further specializing
to the case $q^{1}=e$ and $q^{2}=q^{3}=...q^{d}=0$.

\sxn{``$T$-Duality'' in CS Theory}

Let us first review Buscher's duality argument \cite{bus} for the scalar field
in 1+1 dimensions.  We shall later see that a straightforward generalization
works for the CS theory.

Consider the action
\be
S=\frac{R^{2}}{4\pi}\int _{S^{1}\times {\bf R}}d^{2}x\partial _{\mu}\phi
\partial ^{\mu}\phi \label{or}
\ee
with
$\phi$ identified with $\phi + 2\pi$:
\be
\phi \approx \phi +2\pi.
\label{phieq}
\ee
The translational invariance $\phi \rightarrow \phi +\alpha$ of
(\ref{or})
can be gauged to
arrive at the action
\be
\tilde{S} =\frac{R^{2}}{4\pi}\int d^{2}x(\partial _{\mu}\phi +W_{\mu})^{2}
\label{org}
\ee
where $W_{\mu}$ transforms according to
\be
W_{\mu} \rightarrow W_{\mu} -\partial _{\mu}\alpha \label{ek}
.\ee

We now introduce a Lagrange multiplier field $\lambda$ which constrains
$W$ in the following way:
\begin{eqnarray}
F\equiv dW&=&0 ,\label{curv}\\
\oint _{C}W &\in &2\pi {\bf Z} .\label{hol}
\end{eqnarray}
Here $C$ is any closed loop, space-like or time-like. [This latter possibility
arises if we identify $t=-\infty$ with $t=+\infty$ in the functional
integral so that the transition amplitude is between an initial state and a
final state obtained after transport around a {\em closed} path in the
configuration space (of fields other than the lagrange multiplier field
$\lambda$).  In this case our manifold
can be thought of as  $T^{2}=S^{1}\times S^{1}$.]  If $W$ satisfies the
conditions
(\ref{curv}) and (\ref{hol}), then it has no observable effects on any other
fields and so it can be ``gauged away'' \cite{bus}
from (\ref{org}) to get back $S$.

Let us thus consider
\be
S'=\frac{R^{2}}{4\pi}\int d^{2}x
(\partial _{\mu}\phi +W_{\mu})^{2} +
\frac{1}{2\pi}\int d\lambda W ,\label{Spr}
\ee
\be
(\partial _{\mu}\phi +W_{\mu})^{2} \equiv
(\partial _{\mu}\phi +W_{\mu}) (\partial ^{\mu}\phi +W^{\mu})
\ee
where $\lambda$ is a function.
It follows from the equations of motion of (\ref{Spr})
that the Lagrange multiplier field $\lambda$ constrains $F(=dW)$ to
be zero.  The condition (\ref{hol}) on the holonomies to be
quantized also follows \cite{lag} if we {\em require} that $\lambda$ be
 identified
with $\lambda +2\pi$ just as $\phi$ itself was. (In the Appendix, we show
 that this latter condition is in fact necessary for the theory to be
consistent).

An alternative derivation of
the quantization of the holonomies uses the functional integral approach
and is as follows. Consider the path integral \be
Z_{\lambda}:=\int {\cal D}\lambda e^{\frac{i}{2\pi}\int d\lambda W }
\label{zlam}
\ee
(which is the part of the full path integral that involves $\lambda$).
Since $\lambda \approx \lambda +2\pi$, we can expand $d\lambda$ according
to
\be
d\lambda =\sum _{n}\alpha _{n}d\lambda _{n}^{(0)} +n_{x}\omega _{x}+n_{t}\omega
_{t},\;\; \alpha _{n}\in {\bf R},\;\; n_{x},n_{t}\in {\bf Z}.\label{explam}
\ee
Here $\lambda _{n}^{(0)}$ is a complete set of single-valued functions on
$T^{2}$ while $\omega _{x}$ and $\omega _{t}$ are one-forms
such that
\be
\oint _{x}\omega _{x}=\oint _{t}\omega _{t}=2\pi ,\label{normomeg}
\ee
\be
\oint _{t}\omega _{x}=\oint _{x}\omega _{t}=0  \nonumber
\ee
where $\oint _{x}$ and $\oint _{t}$ refer respectively to the integrals along
the circles in $x$ and $t$ directions.  Thus
\begin{eqnarray}
Z_{\lambda} &=&\sum _{n_x ,n_t }\int \prod _{n}d\alpha _{n}
e^{\frac{i}{2\pi}[\alpha _{n}\int d\lambda _{n}^{(0)}W+n_x \int \omega _{x}W
+n_t \int \omega _{t}W ]}\nonumber\\
&\sim & \prod _{n}\delta [\int d\lambda _{n}^{(0)}W]\sum _{n_x ,n_t
}e^{\frac{i}{2\pi}[n_x \int \omega _{x}W +n_{t}\int \omega _{t}W ]}.
\label{zlamb}
\end{eqnarray}
Now
\be
\sum _{n}e^{inX} =2\pi \sum _{m}\delta (X-2\pi m).\label{Two}
\ee
Therefore
\be
Z_{\lambda}\sim \delta [dW] \sum _{m_1 ,m_2 }\delta (\int \frac{\omega
_{x}}{2\pi}W -2\pi m_1 )\delta (\int \frac{\omega _{t}}{2\pi}W -2\pi m_2 )
\label{zl}
\ee
(where, to get the first delta functional, we have done a partial integration
of the corresponding term in (\ref{zlamb}) and used the fact that $\lambda
_{n}^{(0)}$ forms a basis).

On using (\ref{normomeg}), we now get
\be
Z_{\lambda}\sim \delta [dW] \sum _{m_1 ,m_2 }\delta (\oint _{t}W -2\pi m_1
)\delta (\oint _{x}W-2\pi m_2 ). \label{zfin}
\ee
Here we have used the fact that
$\omega _x$ ($\omega _t$) can be chosen to be independent of $t$ ($x$) by
adding an exact form.  This addition does not affect the values of the
integrals
in (\ref{zl}) because of the multiplying delta functional $\delta [dW]$.

Thus the conditions (\ref{curv}) and (\ref{hol}) follow.

Under these conditions, we can therefore gauge away $W$ to get back the
original action (\ref{or}).

If on the other hand, we decide to integrate out
the $W$ field first, then
\begin{eqnarray}
Z_{W}&=&\int {\cal D}We^{i[\frac{R^{2}}{4\pi} \int (\partial _{\mu}\phi
+W_{\mu})^{2} +\frac{1}{2\pi}\int \epsilon ^{\mu\nu}\partial _{\mu}\lambda
W_{\nu}]} \nonumber\\
&\sim & e^{i[-\frac{R^{2}}{4\pi}\int (\partial _{\mu}\phi-\frac{1}{R^{2}}
\epsilon ^{\mu\nu}\partial _{\nu}\lambda )^{2} +\frac{R^{2}}{4\pi}\int
(\partial
_{\mu}\phi )^{2}]}\nonumber\\
&\sim & e^{i[\frac{1}{2\pi }\int d\phi d\lambda +\frac{1}{4\pi R^{2}}\int
(\partial _{\mu}\lambda )^{2}]} \nonumber\\
&\sim & e^{\frac{i}{4\pi R^{2}}\int (\partial _{\mu}\lambda )^{2}}.
\label{Three}
\end{eqnarray}
We have used the fact here that $e^{\frac{i}{2\pi}\int d\phi d\lambda}=1$
which is a consequence of the identification
 $\phi \approx \phi +2\pi$ and $\lambda \approx
\lambda +2\pi$.

Thus the theory we get now has the ``dual'' action
\be
S_{d}=\frac{1}{4\pi R^{2}}\int (\partial _{\mu}\lambda )^{2}. \label{daction}
\ee

This completes our review of the duality argument for the scalar field theory.

We will now repeat this argument for the CS case.  To begin with we
have the action
\be
S=\frac{k}{2\pi}\int _{M}\alpha d\alpha ,\label{in}
\ee
$M$ being an oriented three-
manifold with an annulus (say) as its spatial slice,
and with time compactified to a circle.  This latter condition is equivalent
to assuming that the fields at $t=\pm \infty$ take the same values so that the
path integral (restricted to the Lagrange multiplier field that will be
introduced shortly) leads
to the transition amplitude between states after
transport around a closed loop in the configuration space (consisting of fields
other than the Lagrange multiplier field).

 As with the scalar field, here too we need an extra
condition on the $\alpha$'s which disallows {\em arbitrary rescalings} of the
$\alpha$.  Without such a condition, $k$ can be changed to $\lambda
^{2} k $ by changing $\alpha $ according to the scheme
$\alpha \rightarrow \alpha \lambda $  ,  $\lambda $ being a real number.
The condition that we impose is
\be
\oint _{C\in \partial M}\alpha \in 2\pi {\bf Z},\label{topo}
\ee
where $C$ is any closed loop on the boundary $\partial M$ of the manifold.
This condition is to be thought of as a generalization of the condition $\phi
\approx \phi +2\pi$ on the scalar field.

Under the transformation
\be
\alpha \rightarrow \alpha + \omega   \label{nonumber}
\ee
on $\alpha$ where $\omega $ is a closed one-form, the Lagrangean three-form
is not invariant, but changes by an exact three-form:
\be
\alpha d \alpha \rightarrow \alpha d \alpha - d ( \omega \alpha )
\nonumber
\ee
We can make it exactly invariant by introducing a ``connection"
one-form $A$, transforming according to
\be
A \rightarrow A - \omega  \label{trA}
\ee
and ``gauging'' $S$ to obtain
\be
\tilde{S}=\frac{k}{2\pi}\int \alpha d\alpha +\frac{k}{2\pi}\int Ad\alpha .
\label{ing}
\ee

But the action $\hat{S}$ is obviously not equivalent to the action $S$ because
the equations of motion are different.
We therefore introduce a Lagrange multiplier one-form $\lambda$ as before to
constrain $A$ by the equations
\begin{eqnarray}
dA&=&0,\label{curva}\\
\oint _{C\in \partial M}A&\in &2\pi {\bf Z}\label{holo}.
\end{eqnarray}
When $A$ fulfills (\ref{curva}) and
(\ref{holo}), we can redefine $\alpha$
using the transformation (\ref{nonumber}) and get back (\ref{in})
and (\ref{topo}).

We thus write
\be
S'=\frac{k}{2\pi}\int \alpha d\alpha +\frac{k}{2\pi}\int Ad\alpha
+\frac{1}{2\pi}\int d\lambda A,\label{ingl}
\ee
where
\be \label{Four}
\oint _{C\in \partial M}\lambda \in 2\pi {\bf Z}.
\ee

Consider
\be
Z_{\lambda}=\int {\cal D}\lambda e^{\frac{i}{2\pi}\int d\lambda A}. \label{Zla}
\ee
to see how (\ref{curva}) and (\ref{holo}) emerge when we integrate out
$\lambda$.
If now, each connected component of the boundary $\partial M$ of $M$, denoted
by $(\partial M)_a$, contains $p_a$  cycles
$C_{ai}$ which can serve to define the generators of its first homology group,
then there exist also $p_a$ closed
one-forms $\omega _{ai}$ (for each $a$) on $\partial M$ such that
\be
\oint _{C_{aj}}\omega _{a'i}=2\pi \delta _{ij} \delta _{aa'}\;\; ,\;
i,j=1,2,\ldots ,p_a .\label{omegas}
\ee
[We assume that the above homology group is torsion-free.]
If $M$ is compact, as we assume,
$(\partial M)_a$ is compact and has no boundaries.  As $M$
is oriented, $\partial M$ too is oriented.  Hence each connected component
$(\partial M)_a$ is a sphere with handles and its homology group has an even
number of generators \cite{?}.  [When the spatial slice is an
annulus (say), $\partial M$ is $T^2 \sqcup T^2$.]  Hence $p_a$
has to be even. In this case we can order the $\omega _{ai}$'s such that
\cite{?}
\be
\int _{\partial M}\omega _{a,2l-1}\omega _{a'j}=4\pi ^{2}\delta _{2l,j} \delta
_{aa'} \;\;\; l=1,2,\ldots ,\frac{p_a}{2}.\label{twod}
\ee
Given any such $\omega$ on $\partial M$, we can associate an $\omega$ on $M$
by requiring \cite{cour}
\be
\nabla ^{2}\omega =0. \label{harm}
\ee
Here $\nabla ^{2}$ is the Laplacian operator on one-forms on
$M$ defined using some Euclidean metric on $M$. [The pull-back of this $\omega$
to $\partial M$ must of course agree with the $\omega$ given there.]
A choice of $\omega _{\bot}$ (the component of $\omega$ perpendicular to
$\partial M$) needs to be made for solving (\ref{harm}).
We can choose it to be zero.

Now, using (\ref{Four}), we can write
\be
\lambda =\lambda ^{(0)}+\sum _{a,i}n_{ai}\omega _{ai}\;\;\; , n_{ai}\in {\bf Z}
\label{dec}
\ee
where $\lambda ^{(0)}$ is a one-form (on $M$) such that
\be
\oint _{C_{aj}}\lambda ^{(0)}=0.\label{lambze}
\ee

Now, given any three-manifold $M$, the operator $*d*d$ (defined by
choosing some Euclidean metric on $M$) on the space of one-forms $\gamma$
admits the following boundary condition compatible with the
self-adjointness of
$*d*d$ (the inner product being defined using the same Euclidean metric)
\cite{mcs}:
\be
\mbox{Pull-back of } \gamma \mbox{ to }\partial M \equiv
\gamma |_{\partial M}=0. \label{sae}
\ee
This means that the one-form $\lambda ^{(0)}$ of equation (\ref{dec}) can be
expanded in a basis of one-forms $\gamma _{n}$
which satisfy the above boundary condition (as in a Fourier expansion so that
the convergence is only in the ``mean-square'' sense).
  Therefore
\be
\lambda =\sum _{n}\beta _{n}\gamma _{n}+\sum _{a,i}n_{ai}\omega _{ai},
\label{decom}
\;\;\;\;  \gamma _{n} | _{\partial M} =0.
\ee

Thus
\begin{eqnarray}
Z_{\lambda}&=&\sum _{n_{ai}}\int \prod _{n}d\beta _{n}e^{\frac{i}{2\pi}[\sum
_{n}\beta _{n}\int _{M}d\gamma _{n}A +\sum _{a,i}n_{ai}\int _{M}d\omega
_{ai}A}]\nonumber\\
&\sim &\prod _{n}\delta (\int d\gamma _{n}A)\sum _{n_{ai}}e^{\frac{i}{2\pi}\sum
_{a,i}n_{ai}\int _{M}d\omega _{ai}A} \nonumber\\
&\sim & \delta [dA] \sum _{n_{ai}}e^{\frac{i}{2\pi}\sum _{a,i}n_{ai}\int
_{\partial M}\omega _{ai}A}. \label{simpli}
\end{eqnarray}
[In arriving at the delta functional here, we have done a partial integration
and
used the completeness of the $\gamma _{n}$'s while to
arrive at the integral in the exponent, we have again done a partial
integration and then neglected the bulk term.  The latter is justified
owing to the multiplying delta functional.]

As before (see (\ref{Two})), this means that
\be
Z_{\lambda}\sim \delta [dA] \prod _{a,i}(\sum _{m_{ai}}\delta
(\int _{\partial
M}\frac{\omega _{ai}}{2\pi}A-2\pi m_{ai})),\;\;\; m_{ai}\in {\bf Z}.
\label{Zfi}
\ee
Since the delta functional above implies that $A$ is a closed one-form, we
can expand $A$ on the boundary $\partial M$ as
\be
A|_{\partial M}=d\xi +\sum _{a,i}r_{ai}\omega _{ai} \label{Five}
\ee
where $\xi$ is a function on $\partial M$ and $r_{ai}$ are valued in reals.

Substituting (\ref{Five}) in the second delta function in (\ref{Zfi}), and
using
(\ref{twod}) and the fact that $\int _{\partial M}\omega _{ai}d\xi =0$ (
$\omega _{ai}$'s being closed one-forms {\em at the boundary}), we
finally get
\be
Z_{\lambda}\sim \delta [dA]\prod _{a,i}(\sum _{m_{ai}}\delta (r_{ai}-m_{ai})).
\label{ZFi}
\ee
Thus, integrating out $\lambda$ gives exactly the conditions (\ref{curva}) and
(\ref{holo}) that we wanted and shows that $S$ is equivalent to the
original action (\ref{in}).

If on the contrary, we choose to integrate out the $A$ field from the action
$S'$ in (\ref{ingl}), we get
\begin{eqnarray}
Z_{A}&=&\int {\cal D}Ae^{i[\frac{k}{2\pi}\int \alpha d\alpha
+\frac{k}{2\pi}\int
Ad\alpha +\frac{1}{2\pi}\int d\lambda A]}\nonumber\\
&\sim &\delta (\frac{k}{2\pi}d\alpha -\frac{1}{2\pi}d\lambda
)e^{i\frac{k}{2\pi}\int \alpha d\alpha }.\label{zforn}
\end{eqnarray}
Since the delta functional here implies that
\be
d\alpha =\frac{1}{k}d\lambda ,\label{condn}
\ee
we have
\be
\alpha =\frac{1}{k}\lambda +\omega ^{(1)} ,\label{stoke}
\ee
$\omega ^{(1)}$ being a closed one-form on $M$.

Thus $Z_{A}$ can be simplified to
\be
Z_{A}\sim \delta (\frac{k}{2\pi}d\alpha -\frac{1}{2\pi}d\lambda )e^{\frac{i}{2
\pi k}\int _{M}\lambda d\lambda +\frac{i}{2\pi}\int _{M}\omega ^{(1)}d\lambda
}\label{simplif}
\ee
The last term in the exponent above is a surface term because $\omega
^{(1)}$ is a closed one-form.
Hence the ``dual" action obtained by integrating out $A$ is
\be
S_{d}=\frac{1}{2\pi k}\int _{M}\lambda d\lambda -\frac{1}{2\pi}\int _{\partial
M}\omega ^{(1)}\lambda \label{Dual}
\ee
where $\lambda $ is subject to the condition
\be
 \oint _{C\in \partial M}\lambda \in 2\pi {\bf Z}, \;\; C=\mbox{ any cycle on
}\partial M.
\ee

Since the second term in (\ref{Dual}) is a surface term, it does not contribute
to the equations of motion.  Moreover, on using the equations of motion
$d\lambda =0$ arising from the first term, we see that the second term
vanishes.

Although we have worked in this Section first with a single scalar field and
then with a single CS field, these considerations generalize (in a sense to be
made precise below) to the case with many scalar fields coupled by a
matrix $G_{ij}$ (as in \cite{dhe}) and the case with many CS fields coupled by
a
matrix $K^{IJ}$ as in the previous Sections.

The procedure to get the dual theory
is always as follows \cite{lag}:

(1) Introduce a gauge field $A$ for some particular transformation
 that is a ``symmetry" of the action
(be it the scalar field theory or the CS theory).  (2) Introduce a
Lagrange multiplier field which constrains $A$ by the conditions
 $dA=0$ and $\oint _{C}A\in 2\pi
{\bf Z}$.  (3) Integrate out the original
 gauge fields to obtain the ``dual'' theory containing the Lagrange
multiplier field.

We get different dual
theories,
depending on the ``symmetries'' we choose to gauge.  It should however be noted
that the duality group that we get using this procedure is still not the full
$O(d,d;{\bf Z})$ \cite{dua} group, because we do not have a method of
incorporating antisymmetric matrices (which are needed for the $O(d,d;{\bf Z})$
transformations) in this approach.

\sxn{Concluding Remarks}

There is a prevalent point of view that
 CS theories involving several vector
potentials are quite effective in reproducing the Hall effect.  In
this paper we have provided further evidence in support of this
viewpoint by showing that
the Haldane hierarchy can be implemented using a sequence of
such CS theories.

The connection of these CS theories to chiral scalar field theories at the edge
has also been demonstrated.  The argument consisted of three stages.  The first
is that the algebra of observables is the same for these two theories.  The
second is that both give rise to the same Hall conductivity in the bulk.  The
third (and perhaps the most important part of the argument) is due to the
fundamental requirement of gauge invariance.  The CS theory when gauged gives
rise to an effective CS theory for the electromagnetic potential.  This is not
gauge invariant and requires a surface action to restore gauge invariance.  The
gauged chiral scalar field theory at the boundary serves precisely as this
surface action.

Another interesting result we have in this paper has to do with a
generalization of the duality transformations for scalar field theories
\cite{dua,bus}.  In a previous work \cite{dhe}, we showed how such duality
transformations relate Hall conductivities at different fractions.  There we
worked purely with a chiral scalar field theory at the edge to arrive at
this result.   It is therefore satisfying to note that analogous duality
transformations exist also for the CS theory in the bulk.  At this point, we
have only an analogue of the $R\rightarrow 1/R$ duality for the CS theory.  It
would be interesting to check if we can also obtain an $O(d,d;{\bf Z})$ duality
for the CS theory with $d$ CS fields.

\centerline{ {\bf Acknowledgements}}

\nopagebreak
We thank T.R. Govindarajan, V. John, G. Jungman, A. Momen and S. Vaidya for
several discussions.  The work of A.P.B. and L.C. was supported by a grant from
DOE, USA under contract number DE-FG02-85ER40231.  The work of L.C. was
supported also by the DOE grant DE-FG05-86ER-40272.

\appendix
\section{Appendix}

In Section 4, the identification of $\lambda$ in (\ref{Spr}) with
$\lambda +2\pi$ was necessary for the arguments presented there.
Here we will give an argument showing that this condition is a
particular case of a more general condition that is required for consistency of
the theory.

Consider the action (\ref{Spr}) rewritten as below:
\be
S' =\frac{R^{2}}{4\pi}\int d^{2}x[(\dot{\phi}+W_0 )^2 -(\phi '+W_1 )^2 ]
+\frac{1}{2\pi}\int d^2 x(\dot{\lambda}W_1 -\lambda 'W_0 ) \label{*}
\ee
Let $\Pi _{\phi}$ be the momentum conjugate to $\phi$.  Then we have the equal
time commutation relation
\be
{[} \Pi _{\phi} (x), \phi (y)]=-i\delta (x-y)
\ee
Also (\ref{*}) implies that $W_{1}$ is the momentum conjugate to $\lambda$ so
that ${[} W_1 (x),\lambda (y)]=-\frac{i}{2\pi}\delta (x-y)$.

The Hamiltonian that follows from (\ref{*}) is
\be
H=\int dx[\frac{\pi}{R^2}\Pi _{\phi}^2 +\frac{R^2}{4\pi}(\phi '+W_1 )^2 ],
\label{**}
\ee
while the Gauss law is
\be
\Pi _{\phi} (x)-\frac{1}{2\pi}\lambda '(x) \approx 0 .\label{***}
\ee
The global Gauss law that follows by integrating above expression over the
whole of space (here a circle) is
\be
Q=\int dx \Pi _{\phi} (x) -\frac{1}{2\pi}\Delta \lambda \;\;\;\;\Delta \lambda
=\lambda (2\pi )-\lambda (0). \label{*4}
\ee

Now, since $\phi \approx \phi +2\pi$,
\be
\mbox{Spectrum of }\int dx \Pi _{\phi} (x) = {\bf Z} +c,\;\;\; c \mbox{ a
constant}. \label{*5}
\ee
As $Q$ must vanish on states, we end up with the requirement
\be
\Delta \lambda =2\pi {\bf Z} +2\pi c. \label{*6}
\ee
With these conditions, $Q$ generates $U(1)$ on arbitrary states, and Gauss law
picks out singlet.

We thus see that the requirement that $\lambda$ be identified with $\lambda
+2\pi$ is natural provided $c$ in (\ref{*5}) vanishes.

Thus for $c =0$, the $\lambda $ obtained from canonical quantization has
an expansion in the form (4.33) showing that the identification of
$\lambda$ with $\lambda + 2\pi $ is natural, in the canonical approach,
for this particular value of
$c$.

Nonzero values of $c$ too can be incorporated in the functional integral
as well by identifying $\lambda$ with $\lambda + 2 \pi (1+c)$.  It
then becomes appropriate for a scalar field canonically quantized with
$c \neq 0$.  But we will not enter into this generalization of the
functional integral here.

\end{document}